\documentclass[fleqn,usenatbib]{mnras}

\usepackage{newtxtext,newtxmath}
\usepackage[T1]{fontenc}

\DeclareRobustCommand{\VAN}[3]{#2}
\let\VANthebibliography\thebibliography
\def\thebibliography{\DeclareRobustCommand{\VAN}[3]{##3}\VANthebibliography}

\usepackage{graphicx}	% Including figure files
\def\gsim{\mathrel{\rlap{\lower 4pt \hbox{\hskip 1pt $\sim$}}\raise 1pt
\hbox {$>$}}}
\def\lsim{\mathrel{\rlap{\lower 4pt \hbox{\hskip 1pt $\sim$}}\raise 1pt
\hbox {$<$}}}

\title[Engine-Driven Supernovae]{Diagnosing The Ejecta Properties of Engine-Driven Supernovae from Observables in Their Initial Phase}

\author[K. Maeda et al.]{
Keiichi Maeda,$^{1}$\thanks{E-mail: keiichi.maeda@kusastro.kyoto-u.ac.jp}
Akihiro Suzuki,$^{2}$ 
and Luca Izzo$^{3}$
\\
$^{1}$Department of Astronomy, Kyoto University, Kitashirakawa-Oiwake-cho, Sakyo-ku, Kyoto, 606-8502. Japan\\
$^{2}$Research Center for the Early Universe, Graduate School of Science, The University of Tokyo, 7-3-1 Hongo, Bunkyo-ku, Tokyo 113-0033, Japan\\
$^{3}$DARK, Niels Bohr Institute, University of Copenhagen, Jagtvej 128, 2200 Copenhagen, Denmark\\
}

\date{Accepted XXX. Received YYY; in original form ZZZ}

\pubyear{2022}

\begin{document}
\label{firstpage}
\pagerange{\pageref{firstpage}--\pageref{lastpage}}
\maketitle

\begin{abstract}
Engine-driven explosions with continuous energy input from the central system have been suggested for supernovae (SNe) associated with a Gamma-Ray Burst (GRB), super-luminous SNe (SLSNe), and at least a fraction of broad-lined SNe Ic (SNe Ic-BL) even without an associated GRB. In the present work, we investigate observational consequences in this scenario, focusing on the case where the energy injection is sufficiently brief, which has been suggested for GRB-SNe. We construct a simplified, spherical ejecta model sequence taking into account the major effects of the central engine; composition mixing, density structure, and the outermost ejecta velocity. Unlike most of the previous works for GRB-SNe, we solve the formation of the photosphere self-consistently, with which we can predict the photometric and spectroscopic observables. We find that these ejecta properties strongly affect their observational appearance in the initial phase ($\lsim$ a week since the explosion), highlighted by blended lines suffering from higher-velocity absorptions for the flatter density distribution and/or higher outermost ejeca velocity. This behaviour also affects the multi-band light curves in a non-monotonic way. Prompt follow-up observations starting immediately after the explosion thus provides key diagnostics to unveil the nature of the central engine behind GRB-SNe and SNe Ic-BL. For SN 2017iuk associated with GRB 171205A these diagnosing observational data are available, and we show that the expected structure from the engine-driven explosion, i.e., a flat power-law density structure extending up to $\gsim 100,000$ km s$^{-1}$, can explain the observed spectral evolution reasonably well. 
\end{abstract}


%%%%%%%%%%%%%%%%%%%%%%%%%%%%%%%%%%%%%%%%%%%%%%%%%%

%%%%%%%%%%%%%%%%% BODY OF PAPER %%%%%%%%%%%%%%%%%%

\section{Introduction} \label{sec:intro}
Massive stars can lead to various explosive phenomena as their end products, following the collapse of the central core. In most of the cases, the outcome is believed to be a core-collapse supernova (CCSN), which is divided into several subclasses based on spectral line footprints \citep[e.g.,][]{filippenko1997}; type II (H-rich), IIb (H-poor), Ib (H-deficient and He-rich), Ic (H- and He-deficient), forming a sequence of envelope stripping during their pre-SN evolution \citep[a red-supergiant progenitor for SNe II to a Wolf-Rayet-like C+O star progenitor for SNe Ic; e.g.,][]{langer2012,maeda2022a}. While the CCSN explosion mechanism has not yet been fully understood, it is widely believed that a neutron star (NS) is formed behind the so-called delayed neutrino-heating mechanism in most, if not all, of these canonical CCSNe \citep[e.g.,][]{janka2012,burrows2021,sato2021}. 

However, more extreme explosions that are probably associated with demise of a massive star have also been found. Gamma-Ray Bursts (GRBs), as a brief burst of gamma-rays \citep[see, e.g.,][for a review]{piran2004}, were proposed to belong to this category \citep{paczynski1986}, and the link to CCSNe was directly confirmed by the discovery of a peculiar SN associated with a weak and nearby GRB 980425 \citep{galama1998}. The SN, termed SN 1998bw, showed much broader spectral features than canonical SNe. A high velocity at the line forming region of C+O-rich ejecta was introduced, which results in a blend of various spectral lines based on SN Ic spectra \citep{iwamoto1998}. Combined with a slower evolution than canonical SNe Ic, it has been proposed that the kinetic energy and the ejecta mass are $\gsim 3 \times 10^{52}$ erg and $\sim 10 M_\odot$, which are both much larger than those derived for canonical SNe Ic \citep{iwamoto1998,woosley1999,nakamura2001,maeda2006}. It has therefore been suggested that SN 1998bw was a hyper-energetic SN Ic following a collapse of a massive star whose main-sequence mass was larger than those triggering canonical CCSNe. 

Since then, an increasing number of SNe have been found to be associated with nearby GRBs when such search for the SN component is possible \citep[but see][]{dellavalle2006,gal-yam2006}, forming `GRB-SNe' \citep[e.g.,][]{woosley2006,hjorth2013,cano2017}. They share the characteristic broad features with the prototypical GRB-SN 1998bw, and thus classified as `broad-lined SNe Ic (SNe Ic-BL)'. Indeed, SNe Ic-BL have been found even without an associated GRB, while the GRB-SNe represent the most extreme cases among SNe Ic-BL \citep{cano2013,modjaz2016}. 

Given the distinct nature of the ejecta properties and the association with a GRB, the explosion mechanism of GRB-SNe (and at least a fraction of SNe Ic-BL) is believed to be different from that behind canonical CCSNe. Given its hyper-energetic nature, it is called the central engine. A popular model is the so-called collapser scenario, which involves a formation of a rapidly-rotating black hole (BH) and an accretion disc in the centre of a collapsing star \citep{woosley1993,macfadyen1999,kumar2008,hayakawa2018}. Another popular model is a `magnetar'-driven scenario, i.e., a formation of a highly-magnetized and rapidly-spinning NS following the core collapse \citep{duncan1992,usov1992,mazzali2006,maeda2007b,metzger2011}. The engine-driven explosion has been further extended as a possible scenario of another flavor of exotic SNe, the H-poor class of `super-luminous SNe (SLSNe-I)' which are characterized by high luminosities and large energy budget in the radiation output \citep{quimby2011,gal-yam2012}. The central engine, similar to the case for GRB-SNe, has been proposed as a power source; the magnetar scenario (\citealt{kasen2010}; see also \citealt{maeda2007a}) or the BH accretion scenario \citep{dexter2013}. 

The continuous energy input from the central engine may change the ejecta dynamics and the composition structure as compared to the canonical SN ejecta. The mixing is an inevitable consequence, which has been found to lead an overall flatter density distribution throughout the ejecta toward the outermost region \citep[with a general structure of $\rho \propto v^{-5}$ or $v^{-6}$ where $\rho$ is the ejecta density as a function of the velocity, $v$, as derived analytically and numerically;][]{suzuki2017,suzuki2019}, rather than the rapidly steepening structure toward the surface usually assumed for canonical CCSNe (especially for SNe Ic) as a result of a `point-like and instantaneous' explosion \citep{matzner1999,tan2001}. The composition structure is also mixed \citep{suzuki2021}. These features are also found at least qualitatively in a jet-like energy input from the central engine, especially along the jet direction \citep{maeda2003,tominaga2007,eisenberg2022,suzuki2022,pais2023}. The scenario further suggests a link between GRB-SNe and SLSNe \citep{metzger2017}; a key may be the difference in the time duration in which the central engine operates, with short-duration and long-duration systems resulting in GRB-SNe and SLSNe, respectively \citep{suzuki2021}. Depending on the engine-operation duration, the outermost ejecta velocity may also be different \citep{eisenberg2022,suzuki2022,pais2023}, which is further affected by the radiation loss \citep{suzuki2021}. 

Deriving the ejecta structure based on observational data of GRB-SNe thus provides powerful diagnostics on the nature of the central engine. It has been performed based on synthetic spectra (for spherical, one-dimensional models) as compared to spectra of individual GRB-SNe (or SNe Ic-BL) \citep[e.g.,][]{iwamoto1998,mazzali2000,nakamura2001,mazzali2002,mazzali2003,mazzali2006,mazzali2006b,ashall2019,izzo2019,ashall2020,kwok2022}. Such study has been showing that observational properties of GRB-SNe and SNe Ic-BL can be generally explained by an energetic explosion of massive C+O-rich ejecta. Further, some degree of the composition mixing \citep[especially of $^{56}$Ni; e.g.,][]{ashall2019,izzo2019} and a flat density structure at a high velocity \citep[e.g.,][]{mazzali2000,izzo2019} have been inferred. 

These results are based on the assumption that a photosphere is formed at a higher velocity than in canonical SNe Ic -- this is indeed a critical shortcoming in most of the previous works, which parameterized the position (velocity) of the photosphere and its temperature (or luminosity). Surprisingly, the spectral synthesis simulations for GRB-SNe based on the self-consistent calculations for the nature of the photosphere are generally lacking, with only a few exceptions as summarized below. \citet{dessart2017} 
performed detailed non-LTE (Local-Thermodynamic Equilibrium) radiation transfer simulations for 1D spherical models, including the self-consistent photosphere-formation processes. However, their models did not take into account the expected effects of the central engine as mentioned above, resulting in a relatively poor fit to the observed spectra of SN 1998bw. Multi-dimensional radiation transfer simulations were performed by \citet{rapoport2012} based on a specific series of parameterized 2D jet-like explosion models of \citet{maeda2002}. Similarly, \citet{barnes2018} and \citet{shanker2021} performed multi-dimensional radiation transfer based on a series of parameterized jet-like explosion models. While these multi-dimensional studies are regarded as being more realistic than one-dimensional models, a drawback is that the results are relatively limited to the particular models which do not necessarily cover a wide range of the possible effects of the central engine. Most critically, all these works focused on the maximum-phase spectral formation, lacking the spectral-model prediction in the first week since the explosion. 

Indeed, the very early-phase spectra potentially provide powerful diagnostics on the nature of the central engine; the central engine manifests its nature in the outermost ejecta, both in the density and composition structures. Given that the ejecta become progressively transparent from the outer region as time goes by, the spectral-formation site initially resides in the outermost layer while it eventually recedes to the inner, lower velocity region. The power of the infant spectroscopic data was demonstrated by \citet{izzo2019} who inferred the existence of extremely high-velocity materials reaching to $\sim 100,000$ km s$^{-1}$ in SN 2017iuk associated with GRB 171205A, based on spectral-synthesis models for its spectra starting on day 1 since the GRB \citep[see also][for another example, GRB 161219B/SN 2016jca]{ashall2019}. Recently added is SN Ic-BL 2020bvc, which showed similarly high-velocity absorption features reaching to $60,000-70,000$ km s$^{-1}$ on day 1.5, despite its non-association with a GRB \citep{ho2020,izzo2020,rho2021}; the development of high-cadence optical surveys now allows such prompt spectroscopic followup observations even without an associated GRB, which has led to the discovery that some of SNe Ic-BL without a GRB can also exhibit extreme observational properties as demonstrated by SN 2020bvc.

In the present work, we {\em predict} the photometric and spectroscopic properties of the engine-driven SNe including a self-consistent photosphere-formation processes. Our investigation is based on idealized/simplified 1D ejecta models, which however includes key effects of the engine-driven explosion in the density and composition structure. We especially focus on the initial, rising-phase behaviours and investigate how such observations can be used to diagnose the ejecta structure and thus the nature of the central engine. In Section 2, we describe our method and models. The results are presented in Section 3, where photometric and spectral properties are addressed in Sections 3.1 and 3.2, respectively. The paper is closed in Section 4 with concluding remarks.

\section{Method and Models}\label{sec:method}
In the present work, we restrict ourselves to one-dimensional, spherical models, based on the success of such models to reproduce basic light-curve and maximum-phase spectral properties (see references in Section 1). This is mainly to simplify/idealize the problem so that we can extract characteristic and general properties especially associated with the outermost ejecta, but also helps to reduce computational time. This is indeed a good approximation for quasi-spherical energy input from the central engine \citep[e.g.,][]{suzuki2017,suzuki2019}, but also can capture basic ejecta properties of a jet-like energy input at least along the line-of-sight if the associated hydrodynamics effects \citep[e.g.,][]{suzuki2022} are properly parameterized (see below). 

We assume that the ejecta structure is immediately frozen after the shock breakout and possible energy input from the central engine, well before observations. Further, we assume that the thermal energy input is dominated by the $^{56}$Ni/Co radioactive heating; these conditions require that the progenitor radius is sufficiently small (which is usually met for a compact C+O star progenitor) and that the energy input by the central engine is terminated well before the time of observations \citep[$\lsim 10^4$ sec or so;][]{suzuki2021}\footnote{Note that the present model is applicable both to the `jet-driven explosion' \citep[e.g.,][]{suzuki2022} and the `quasi-spherical explosion followed by an additional energy input by the central engine' \citep[e.g.,][]{suzuki2017}.}. When both of these requirements are satisfied, the energy deposited either by the initial shock or subsequent operation of the central engine, or both, should be quickly converted to the kinetic energy, leading to the freezeout of the ejecta structure and the homologous expansion. This is a scenario appropriate for GRB-SNe \citep{suzuki2021}. The opposite case, where the central engine operates in a long time scale and thus determines the ejecta energy content, will be investigated in the future as a scenario for SLSNe (K. Maeda, in prep.).

The properties of the ejecta model in the present work are thus specified by the (one-dimensional) density structure ($\rho (v, t) = \rho_0 (v) t^{-3}$) and the distributions of different elements and radioactive isotopes ($X_{\rm i} (v)$), where the relations are expressed in the velocity space (i.e., the Lagrangian coordinate). For the density structure, our reference model is the CO138 model applied to the prototypical GRB-SN 1998bw, which was constructed by a one-dimensional thermal-bomb hydrodynamic simulation based on a $13.8 M_\odot$ C+O star progenitor model \citep{iwamoto1998,nakamura2001}. The ejecta mass and the kinetic energy are $10.6 M_\odot$ and $3 \times 10^{52}$ ergs, respectively. In addition to the original CO138 density structure, we test a power-law density distribution, e.g., $\rho_0 (v) \propto v^{-6}$; the energy input by a central engine introduces substantial mixing within the ejecta, and the resulting density structure is well described by a power-law distribution with the index of $-5$ to $-6$ as a function of the velocity \citep{suzuki2017,suzuki2019,suzuki2021}. A similar mixing effect is found for a jet-like energy injection along the jet axis \citep{eisenberg2022,suzuki2022,pais2023}, and our one-dimensional simulation may also be regarded as a rough approximation along the jet axis which is expected to be aligned with the line-of-sight for GRB-SNe (see also Section 4 for further discussion). The power-law distribution is constructed by hands, keeping the same ejecta mass and kinetic energy as in the original CO138 model. 

\begin{table}
\centering
\caption{Models: $^{a}$For the density structure, `CO138' adopts the original 1D hydrodynamic model, while `POW' adopts the power-law distribution ($\rho \propto v^{-6}$). For the composition structure, `nomix' adopts the original stratified structure, while `mix' adopts the homogeneous mixing. 
}
\label{tab:models}
\begin{tabular}{cccc} % four columns, alignment for each
\hline
Model$^{a}$ & Density & Mixing & $v_{\rm max}$ ($10^{9}$ cm s$^{-1}$) \\
\hline
 CO138\_nomix\_10 & Original & No & 10\\
 CO138\_mix\_8 & Original & Yes & 8 \\
 CO138\_mix\_10 & Original & Yes & 10 \\
 POW\_mix\_6 & Power & Yes & 6 \\
 POW\_mix\_8 & Power & Yes & 8 \\
 POW\_mix\_10 & Power & Yes & 10 \\
 POW\_mix\_12 & Power & Yes & 12 \\
\hline
\end{tabular}
\end{table}

The operation of the central engine will also introduce the mixing of the chemical compositions \citep{maeda2002,maeda2003,tominaga2007,suzuki2021,suzuki2022}. To examine the effect of the composition structure in the observables, we examine two cases; one adopting the (original) stratified structure, and the other adopting the homogeneous mixing of all the elements and $^{56}$Ni introduced by hands. Given that the power-law density model already assumes a large-scale mixing, we adopt only the mixing case for the power-law model. In both cases, the solar abundance compositions are added to the heavy elements beyond neon. We however note that adding the solar compositions introduces only a minor modification for the mixing model, since the metal content is dominated by the newly-synthesized elements everywhere in the ejecta.

An additional model parameter in our model sequence is the outermost ejecta velocity ($v_{\rm max}$), above which the density is set zero. This quantity depends on the hydrodynamics of the shock propagation in the steeply decreasing progenitor density structure \citep{matzner1999}, together with the energy redistribution following the shock breakout due to the radiation loss \citep{falk1977,ensman1992,suzuki2016}. The operation of the central engine further provides an additional factor, again through the combined effect of the hydrodynamic evolution \citep{suzuki2019,pais2023} and radiation loss \citep{suzuki2021}. In other word, our aim here is to clarify what observables keep the information on the outermost ejecta velocity, and how the observations can be used to constrain it; we thus aim at providing possible diagnostics on the nature of the central engine through $v_{\rm max}$. 

Introducing $v_{\rm max}$ and the power-law distribution is conducted in a way to keep the ejecta mass and the kinetic energy in the original CO138 model ($10.6 M_\odot$ and $3 \times 10^{52}$ ergs), by adjusting/scaling the innermost velocity cut and the density scale, i.e., $\rho_0 (v)$ for given $v$. 
The innermost velocity cut does not affect the present conclusions at all, as the innermost part is not exposed in the time window of interest in the present work. The effect of the density scale is also minimal; even for the two extreme choices of $v_{\rm max} = 120,000$ and $60,000$ km s$^{-1}$ in the power-law density model sequence, the difference in the density scale is only $\sim 15$\% since the mass content in the outermost layer are negligible even for the power-law distribution sequence since it is much steeper than $\rho_0 (v) \propto v^{-3}$. The (homogeneous) mass fractions of different elements are computed to keep the masses of each element in the original CO138 model (after adding the solar compositions), and the differences in the mass fractions between models with different values of $v_{\rm max}$ are negligible for the same reason. 

\begin{figure}
\centering
\includegraphics[width=\columnwidth]{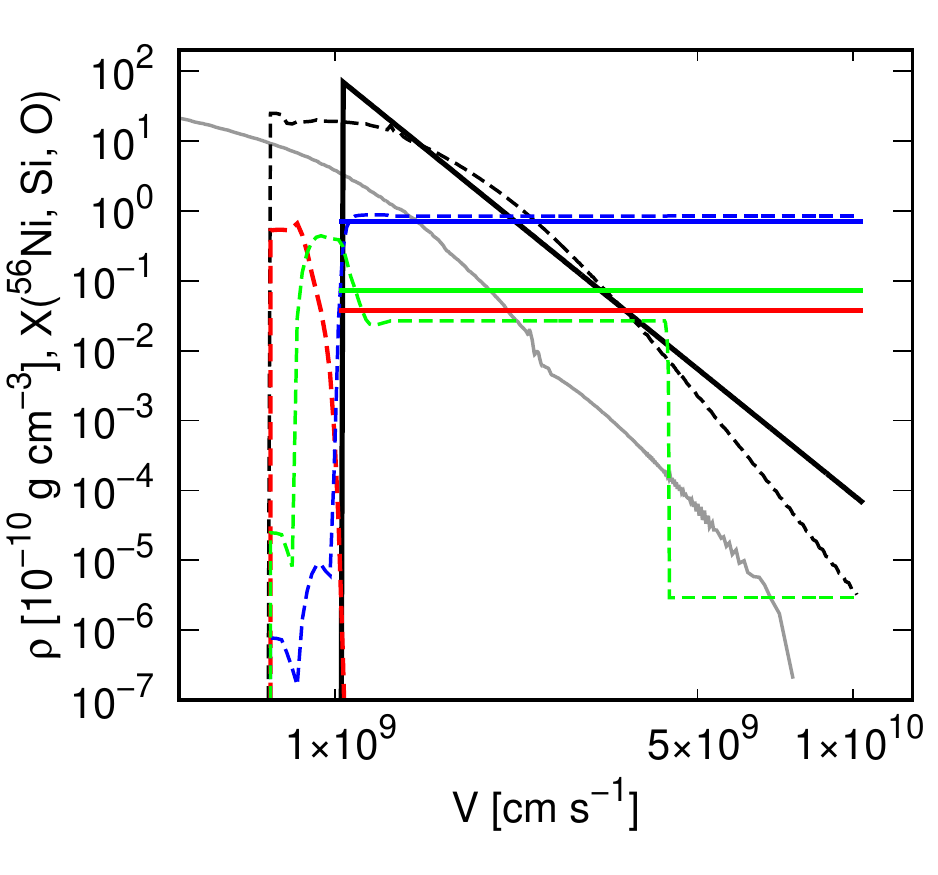}
\caption{The density structure of Models CO138 (black-dashed) and POW (with the power-law index of $-6$; black-solid). The outermost velocity, $v_{\rm max}$, is set as $100,000$ km s$^{-1}$ in this figure. The density is scaled at 1 day since the explosion. The mass fractions of $^{56}$Ni (red), Si (green), and O (blue) are shown, for Models CO138\_nomix\_10 (dashed) and POW\_mix\_10 (solid). The `representative' SN Ic structure taken from \citet{fang2023} is also shown (grey). }
\label{fig:density}
\end{figure}

\begin{figure}
\centering
\includegraphics[width=\columnwidth]{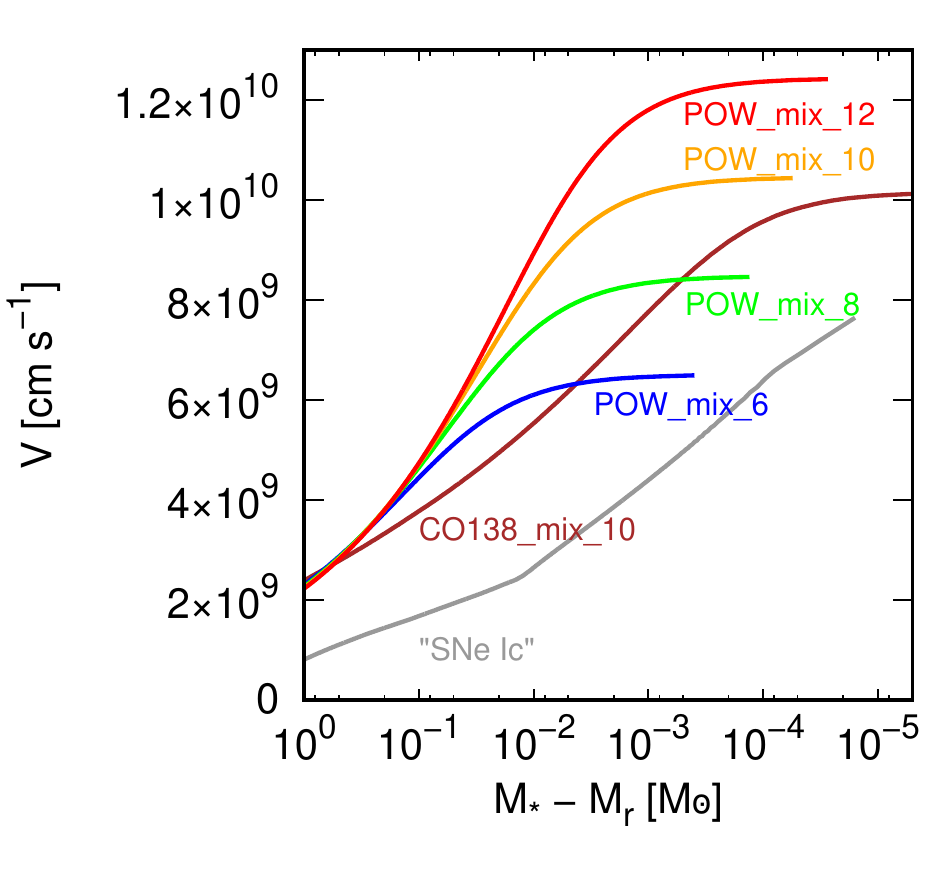}
\caption{The velocity as a function of the mass coordinate as measured from the outermost ejecta. Shown here are the structures of CO138\_mix\_10 (brown), POW\_mix sequence (blue, green, orange, and red for $v_{\rm max} = 60,000$, $80,000$, $100,000$, and $120,000$ km s$^{-1}$, respectively), and the representative SN Ic model (grey).} 
\label{fig:velocity}
\end{figure}

Table \ref{tab:models} summarizes the model sequence examined in the present work. Fig. \ref{fig:density} shows the density structure for Models CO138 and POW, as well as the composition structure for cases `nomix' and `mix'. The mass of $^{56}$Ni is set to be $0.4 M_\odot$ in all the models. To demonstrate the difference in the ejecta structures of the present models and that adopted for a typical SN Ic, we also show a representative `SN Ic' ejecta structure in Fig. \ref{fig:density}; an explosion of a stripped CO core evolved from a $18 M_\odot$ main-sequence star \citep[][their CO18 model]{fang2023}. The ejecta mass and the kinetic energy are set to be the typical values estimated for SNe Ic \citep[$\sim 2.7 M_\odot$ and $2 \times 10^{51}$ ergs;][]{taddia2015,lyman2016}. The differences introduced by the density structure and $v_{\rm max}$ are demonstrated in Fig. \ref{fig:velocity}, which shows the velocities of different layers as a function of the mass contained above a given layer up to the outermost surface of the ejecta. 

\begin{figure*}
\centering
\includegraphics[width=0.68\columnwidth]{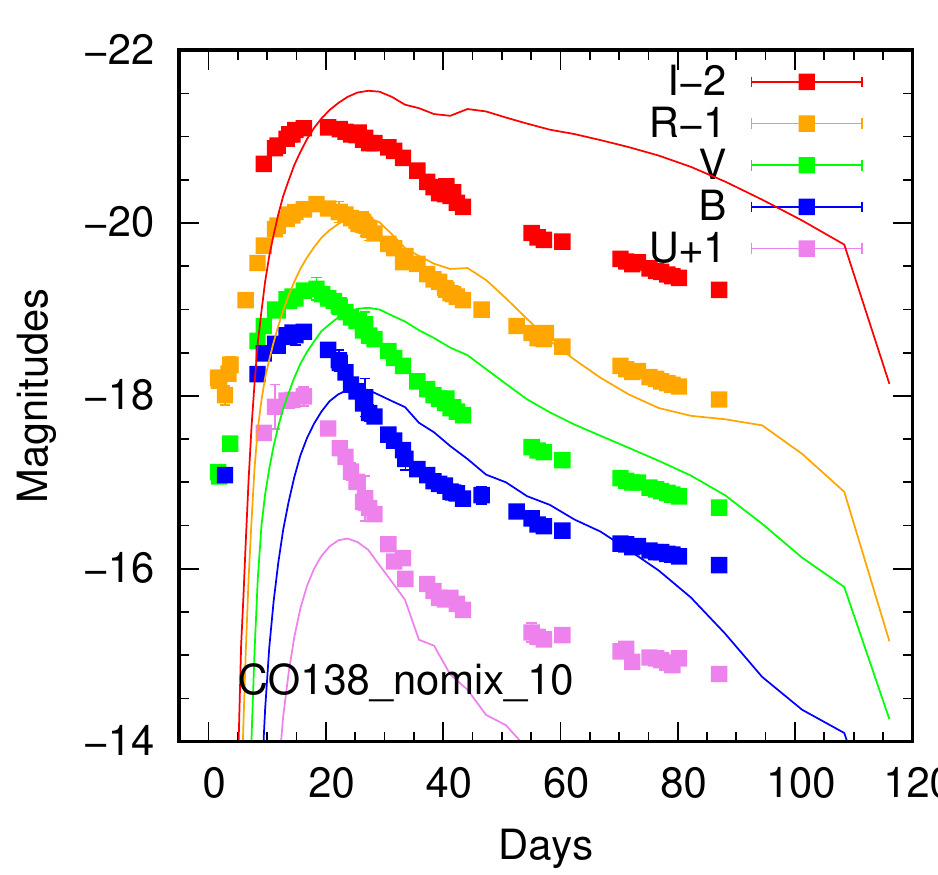}
\includegraphics[width=0.68\columnwidth]{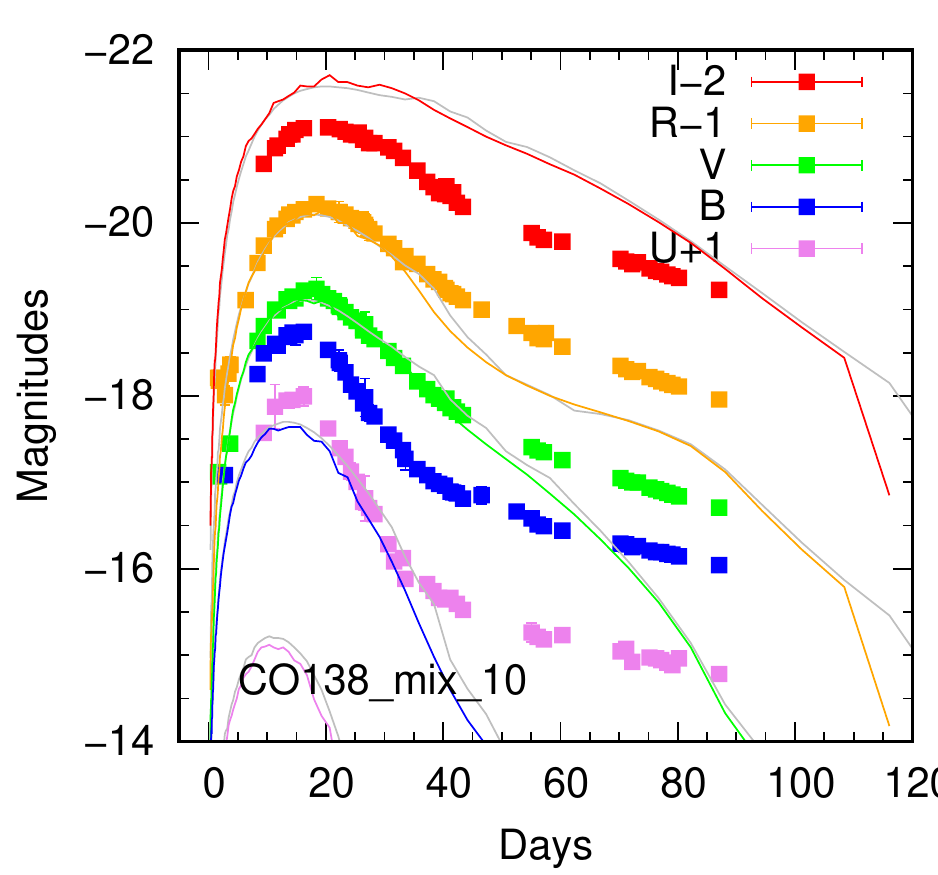}
\includegraphics[width=0.68\columnwidth]{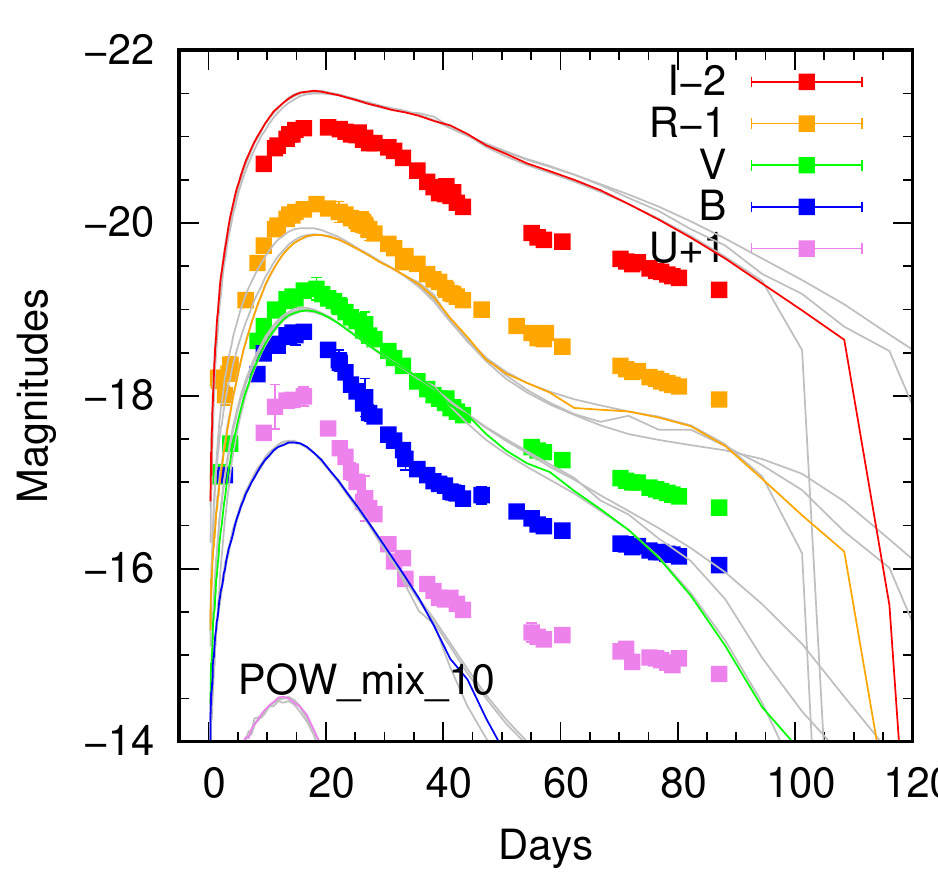}
\caption{The light curves of Models CO138\_nomix\_10 (left), CO138\_mix\_10 (middle), and POW\_mix\_10 (right). The light curves are shown for the $U$, $B$, $V$, $R$, and $I$ bands from bottom to top (shown in violet, blue, green, orange, and red, respectively). The same model sequence with different $v_{\rm max}$ (see Table \ref{tab:models}) is shown by grey lines. For an illustrative purpose, the light curves of SN 1998bw \citep{galama1998,sollerman2002,clocchiatti2011} are also shown by the filled squares, using the same colour coordinate as that used for the models. The Vega system is used throughout the present paper.}
\label{fig:lc}
\end{figure*}

A high velocity could be obtained toward the outermost layer even in the canonical SN Ic ejecta due to the acceleration of the shock wave in propagating the surface of a progenitor star with a steep density gradient \citep{matzner1999}, depending on the radiative loss at the shock breakout \citep{falk1977,ensman1992}. However, in the canonical SN Ic ejecta, the density is very low at the outermost layer (Fig. \ref{fig:density}), and thus changing $v_{\rm max}$ would not introduce much effect on the optical display; the mass of the high-velocity material is too small to participate in the spectral formation (Fig. \ref{fig:velocity}). The CO138 model has a larger amount of materials at a high velocity \citep{iwamoto1998,nakamura2001}, and thus such high-velocity materials potentially affect the radiation transfer, therefore the result can be dependent on the choice of $v_{\rm max}$. Further, the POW model sequence has even a larger amount of materials at a high velocity due to the flat density distribution (Fig. \ref{fig:velocity}); $\gsim 0.04 M_\odot$ and $\gsim 10^{-3} M_\odot$ at $v \gsim 60,000$ and $\gsim 100,000$ km s$^{-1}$, respectively, which are larger than those in the CO138 model ($\sim 5 \times 10^{-3} M_\odot$ at $\gsim 60,000$ km s$^{-1}$ and $\sim 5 \times 10^{-5} M_\odot$ at $\gsim 100,000$ km s$^{-1}$), and much larger than in the `SN Ic' ejecta ($\sim 10^{-4} M_\odot$ at $\gsim 60,000$ km s$^{-1}$). 

For each model structure thus constructed, we simulate multi-band light curves and spectra using HEIMDALL \citep[Handling Emission In Multi-Dimension for spectrAL and Light curve calculations;][]{maeda2006,maeda2014}. Despite the multi-dimensional capability of HEIMDALL, the present investigation is restricted to one-dimensional spherical configuration.

In the radiation transfer simulations, the $\gamma$-ray packets are created through the radioactive input following the $^{56}$Ni distribution given as the model input. The propagation of the high-energy photons originally emitted as the decay lines is followed as a random walk process with the Monte-Carlo method, including the Compton scattering, photoelectric absorption, and pair production as the photon-matter interaction processes \citep{maeda2006b}. The energy deposited to the gas by these processes is recorded, together with the energy provided by positrons immediately following the positron-emission decay channel. 

Once the thermal energy deposition to each spatial and temporal grid is determined, this information is used to create optical photon packets. The transfer of these optical photons is then performed again using the Monte-Carlo method, where iteration is solved in each time step to determine the ionization status, the electron number density, and the temperature, as obtained by requiring the self-consistency between the radiation field and gas conditions. The LTE is assumed. The radiation transfer calculations involve $\sim 5 \times 10^5$ bound-bound transitions, as well as free-free and bound-free transitions \citep{maeda2006}, using the opacity data taken from \citet{kurucz1995}.

The number of the radial grids used in the radiation transfer simulations is $300 - 600$ for the models with $v_{\rm max} = 60,000 - 120,000$ km s$^{-1}$. It then provides the resolution of $< 200$ km s$^{-1}$ in the input ejecta models, which is sufficient to resolve wavelength-dispersion elements in typical spectroscopic observations. The number of the photon packets is $\sim 10^8$, in all the simulations. At the end of each simulation, the photon packets are sampled into 90 bins in time (with a logarithmic spacing between 0.3 and 160 days) and 3,000 bins in wavelengths (with a logarithmic spacing between 100 and 20,000\AA). The number of the photon packets is sufficiently large to achieve the convergence in the thermal condition in each space and time grid, as well as to provide a good Signal-to-Noise ratio in the synthetic spectra (i.e., on average $\sim 300 - 400$ packets in each time-step and wavelength bin).

\section{Results}\label{sec:result}

\subsection{Photometric Properties}

\begin{figure*}
\centering
\includegraphics[width=\columnwidth]{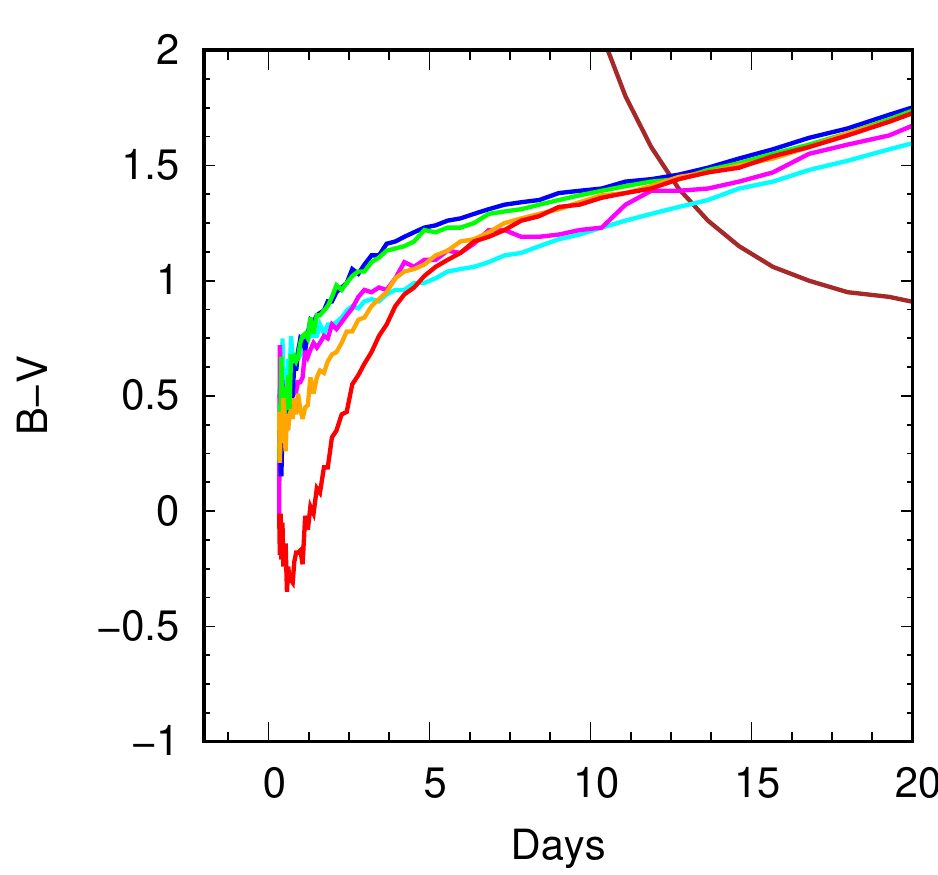}
\includegraphics[width=\columnwidth]{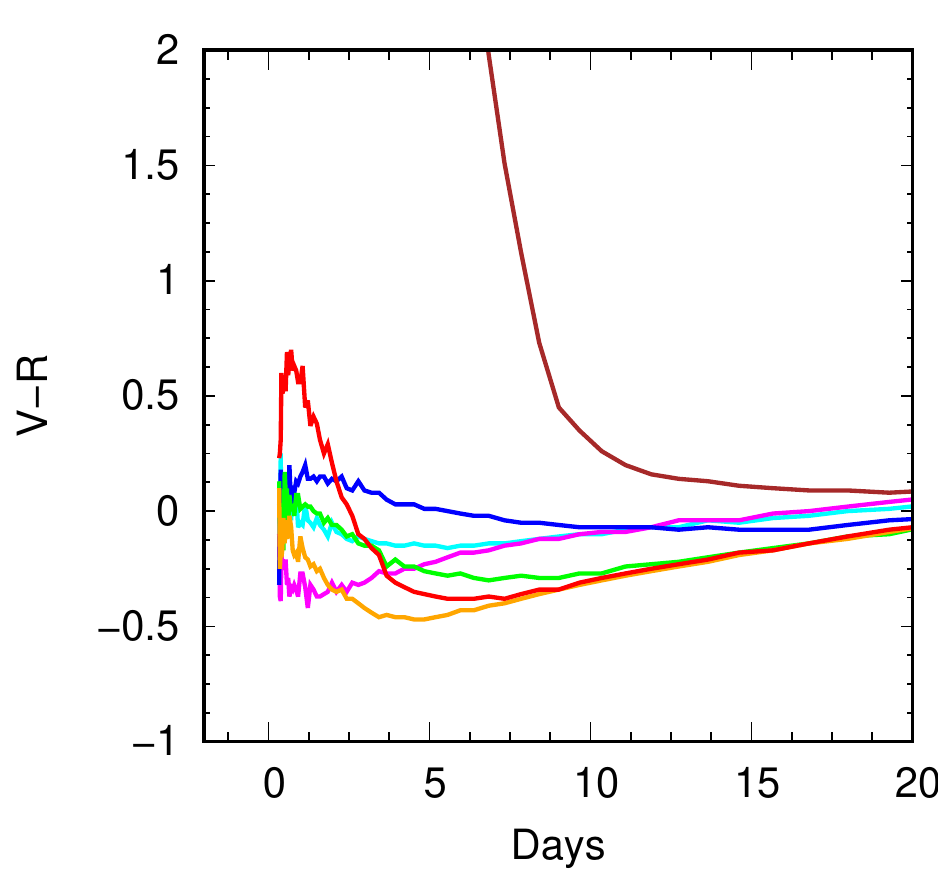}
\caption{The colour evolution of the model sequences CO138\_nomix (brown) and CO138\_mix (cyan and magenta for $v_{\rm max} = 80,000$ and $100,000$ km s$^{-1}$, respectively), and POW\_mix sequence (blue, green, orange, and red for $v_{\rm max} = 60,000$, $80,000$, $100,000$, and $120,000$ km s$^{-1}$, respectively). 
}
\label{fig:color}
\end{figure*}

\begin{figure*}
\centering
\includegraphics[width=\columnwidth]{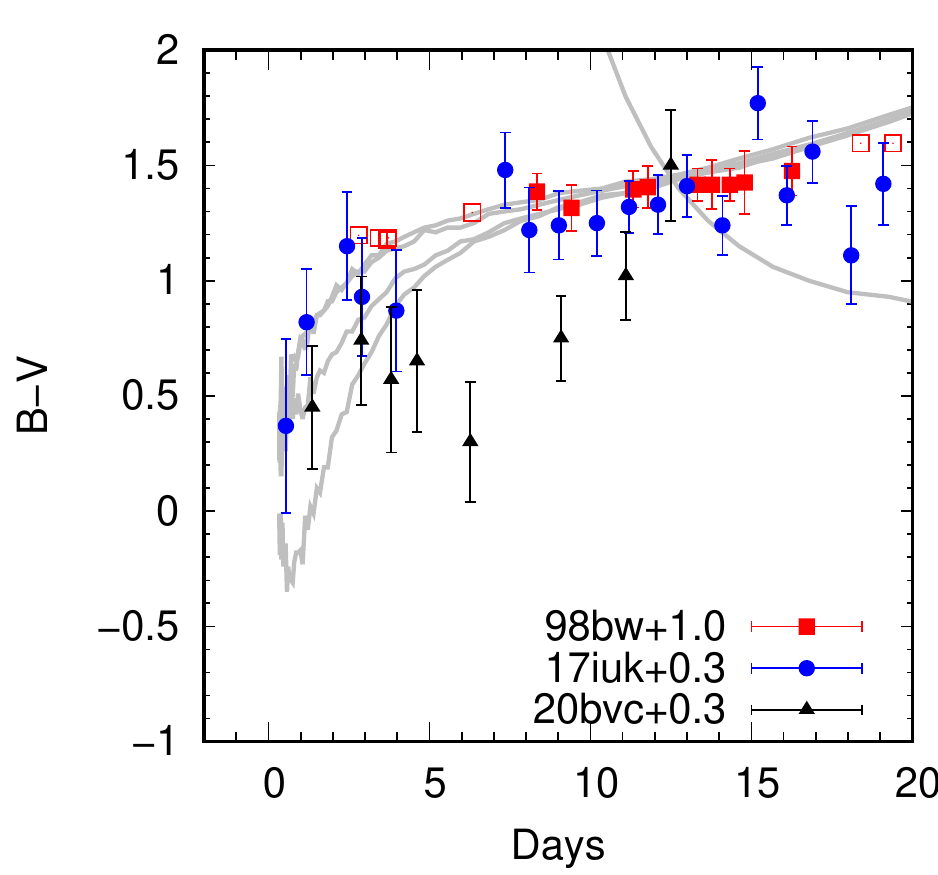}
\includegraphics[width=\columnwidth]{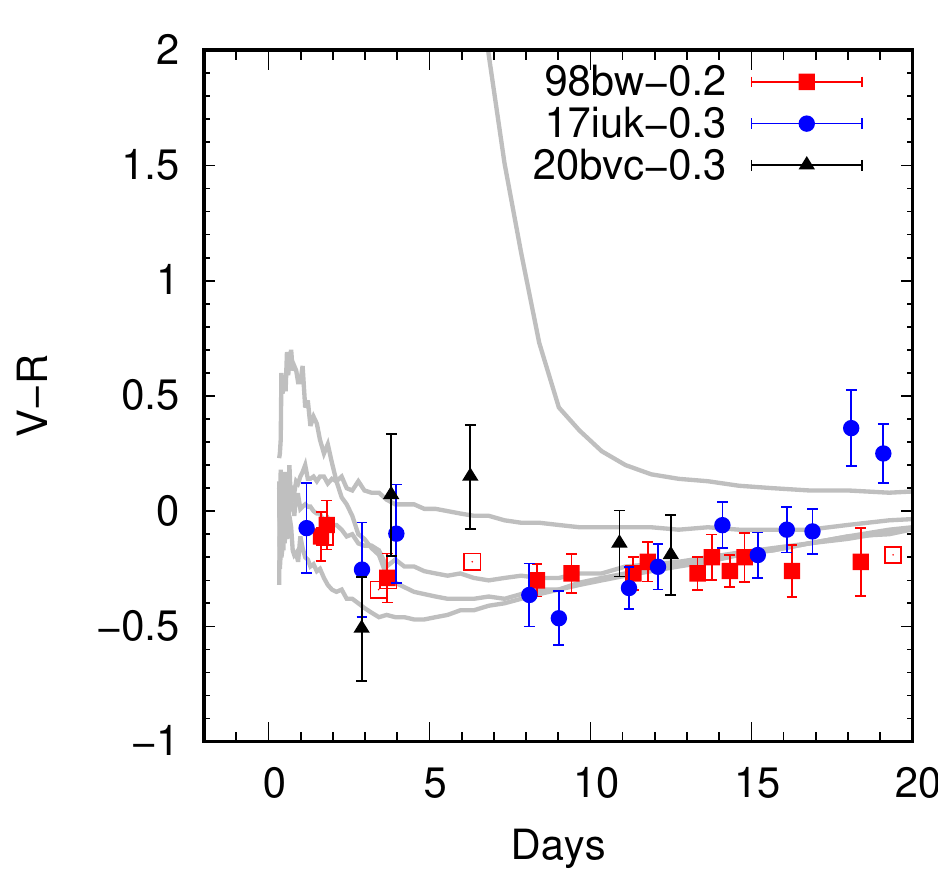}
\caption{The colour evolution of SNe 1998bw (red), 2017iuk (blue), and 2020bvc (black), overplotted with the CO138\_nomix model and the POW\_mix sequence ($v_{\rm max} = 60,000$, $80,000$, $100,000$, and $120,000$ km s$^{-1}$; see Fig. \ref{fig:color}) as shown by grey lines. For SN 1998bw, the colours computed by the magnitudes in the two bands in the close proximity in the observed epochs are shown in the filled squares, while the colours that involve interpolation are shown by open squares. For SNe 2017iuk and 2020bvc, the $B$- and $V$-band magnitudes are obtained with the SWIFT UVOT. For SN 2017iuk, the $V-R$ colour is shown using the data obtained with the GROND $r'$-band filter. For SN 2020bvc, the $R$-band magnitude is replaced by the ZTF $r$-band magnitude. }
\label{fig:color_obs}
\end{figure*}

Fig. \ref{fig:lc} shows the synthetic multi-band light curves of Models CO138\_nomix\_10, CO138\_mix\_10, and POW\_mix\_10. In the same figure, we also plot the multi-band light curves of the prototypical GRB-SN 1998bw, but noting that it is merely for a demonstration purpose since we have not tuned the models to fit to the observational data of SN 1998bw. 

With the mixing of $^{56}$Ni, the light curves show a quicker rise to the peak, and the peak is reached earlier. Accordingly, the $^{56}$Ni and composition mixing affects the colour evolution, as shown in Fig. \ref{fig:color}. These are well-known effects of the extended distribution of $^{56}$Ni \cite[e.g.,][]{nakamura2001,dessart2012,bersten2013,yoon2019}. The need for the chemical mixing, especially of $^{56}$Ni, has indeed been suggested to fit the light curves of SN 1998bw \citep{nakamura2001}, as is evident from Fig. \ref{fig:lc}. The $^{56}$Ni mixing is indeed a property shared by SNe Ic-BL in general \citep{taddia2019}, or even by canonical SNe Ic \citep{taddia2015,yoon2019}. 

We find that the outermost ejecta velocity, $v_{\rm max}$, also affects the initial, rising part of the light curves. This is seen in the colour evolution, as shown in Fig. \ref{fig:color}. As the photosphere recedes in the velocity space, the models with different values of $v_{\rm max}$ converge as the material well above the photosphere starts decoupling from the radiation field. For example, Models POW with $v_{\rm max} = 120,000$ (red) and $100,000$ km s$^{-1}$ (orange) show different colour evolution in the first week. Then these two models converge, but keep showing different colours from Model POW with $v_{\rm max} = 80,000$ km s$^{-1}$ (green) until $\sim 10$ days. After that, these three models become mutually indistinguishable in their colours, but show different colours (especially in the $V-R$ colour) from Model POW with $v_{\rm max} = 60,000$ km s$^{-1}$ (blue). Finally all these models converge to the same colour evolution around the maximum light. 

The effect of the outermost velocity on the $B-V$ colour evolution is monotonic on $v_{\rm max}$ but that of the $V-R$ colour evolution is non-monotonic. These effects are not explained simply by different evolution of the photosphere nor by additional radioactive-decay energy input in the outermost layer for models with larger $v_{\rm max}$. These differences in the colours are indeed driven by the differences in the spectral formation above the photosphere. We will clarify this issue in discussing the spectral formation process in the next section. 

The observed colour evolution of SNe 1998bw, 2017iuk, and 2020bvc is shown in Fig. \ref{fig:color_obs} \citep{clocchiatti2011,izzo2019,izzo2020,ho2020}, overplotted with the CO138\_nomix model and the POW\_mix sequence with different values of $v_{\rm max}$. For SN 1998bw, unfortunately the early evolution is not well sampled; we therefore perform interpolation of the data points to compute the colours (open squares), unless there are data points in the two bands in close proximity in the epochs (filled squares). Fig. \ref{fig:color_obs} is for a demonstration purpose, since the model has not been tuned to individual SNe; as such, direct comparison between the model colour and the observed one should not be conducted, while the colour `evolution' may still be meaningful. 

SN 2020bvc shows a change in the colour evolution at $\sim 5$ days both in the $B-V$ and $V-R$ colours, which is not seen in (or opposite to) the model predictions. Indeed, SN 2020bvc shows the initial rapid decline in the bolometric luminosity, which cannot be attributed to the $^{56}$Ni/Co heating. The present models thus do not apply to the early photometric evolution of SN 2020bvc, and additional powering mechanism must be considered; the issue will be investigated in a forthcoming paper (K. Maeda, in prep.). For the other two SNe, the overall colour evolution is indeed well explained by the present models; especially, it is clear that substantial $^{56}$Ni mixing is necessary to explain the early colour evolution \citep[see also][]{yoon2019}. The initial rapid evolution toward the red in the $B-V$ colour, as well as the transition from the red-to-blue evolution then to the blue-to-red evolution in the $V-R$ colour, are seen in SN 2017iuk as predicted by the models; it is seen that the evolution roughly matches to the POW\_mix models with $v_{\rm max} = 80,000$ to $100,000$ km s$^{-1}$. This is indeed consistent with the conclusion from the spectral modeling \citep[Section 3.2; see also][]{izzo2019}. The photometric data of SN 1998bw are less constraining, due to the sparsely sampled data points; the colours estimated with the interpolation involves large uncertainties and should not be over-interpreted. Still, SN 1998bw shares the qualitative behaviors in the colour evolution with SN 2017iuk and thus with the model predictions. The POW\_mix model with $v_{\rm max} = 80,000$ km s$^{-1}$ provides a better match to the data than the other models (with the above caveat bared in mind). These investigations demonstrate that the early photometric colour evolution can serve as a powerful diagnostics of the outermost ejecta structure, especially the outermost velocity. 

\subsection{Spectroscopic Properties}

\begin{figure*}
\centering
\includegraphics[width=\columnwidth]{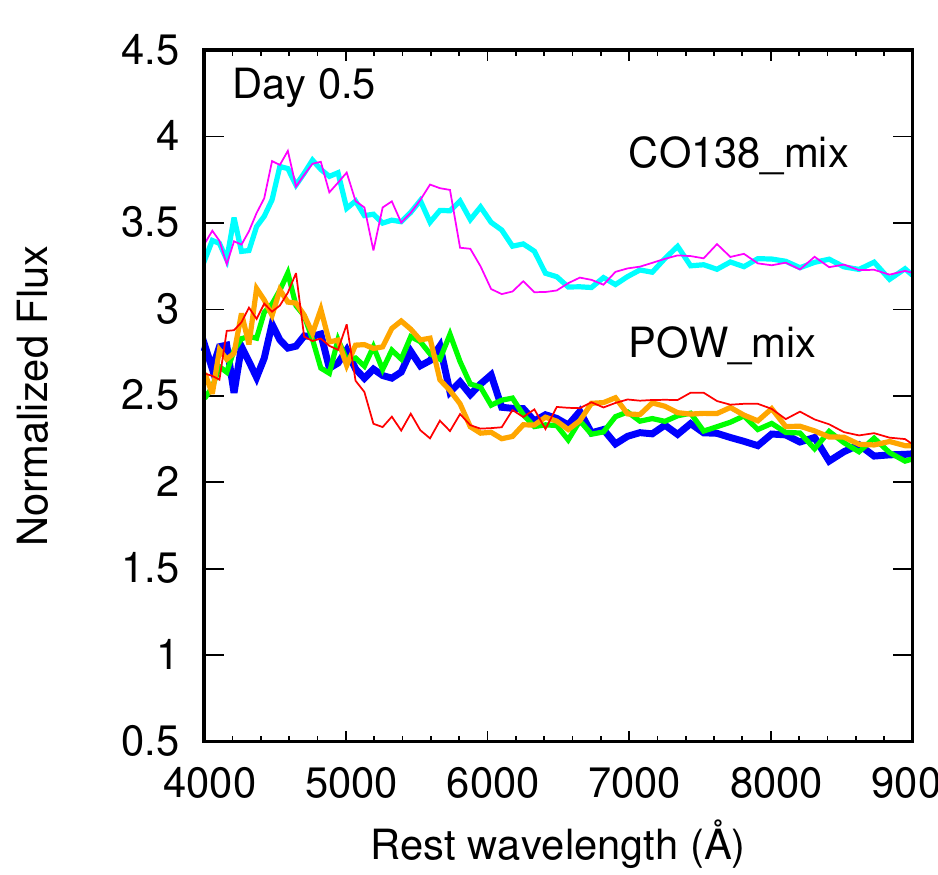}
\includegraphics[width=\columnwidth]{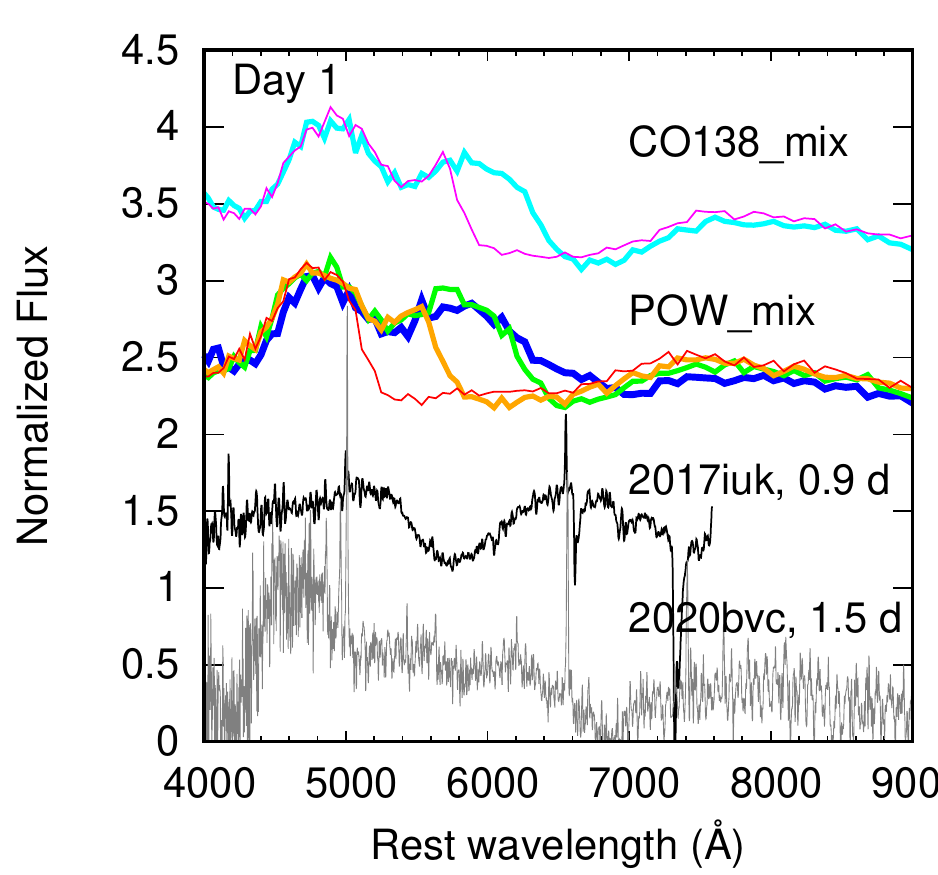}
\includegraphics[width=\columnwidth]{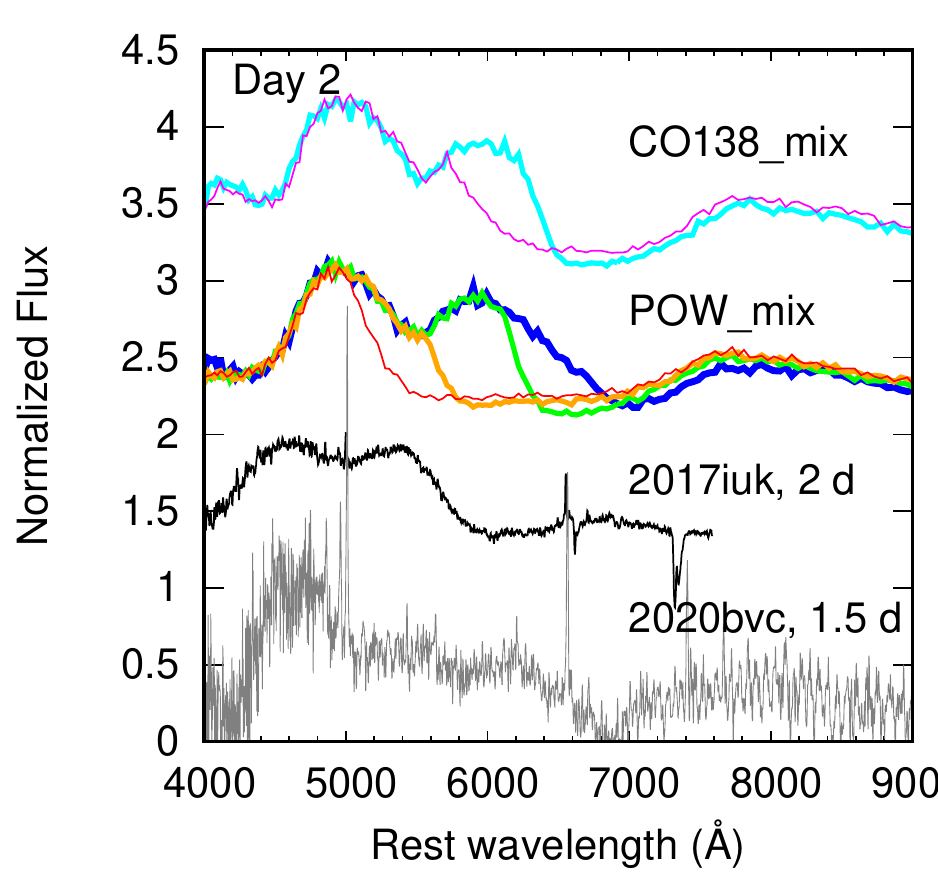}
\includegraphics[width=\columnwidth]{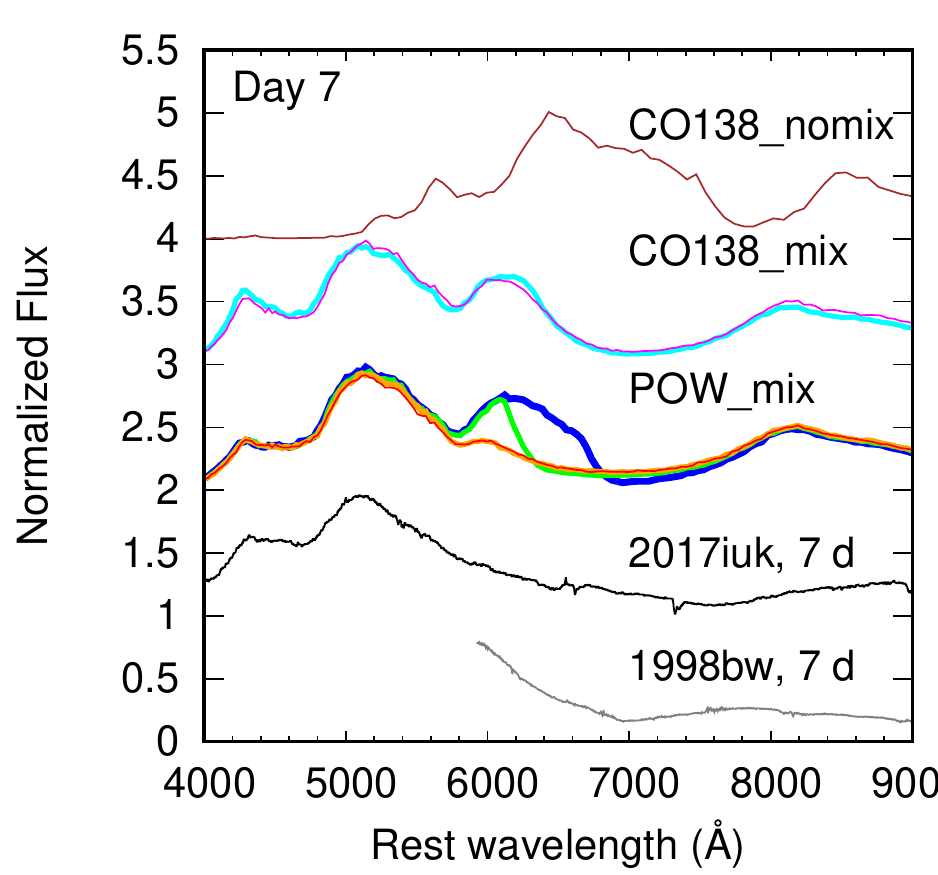}
\caption{The synthetic spectra in the infant to early phases, for Models CO138\_nomix (brown), CO138\_mix with $v_{\rm max} = 80,000$ km s$^{-1}$ (cyan) and $100,000$ km s$^{-1}$ (magenta), POW\_mix with $v_{\rm max} = 60,000$ km s$^{-1}$ (blue), $80,000$ km s$^{-1}$ (green), $100,000$ km s$^{-1}$ (orange), and $120,000$ km s$^{-1}$ (red). Shown here for a demonstration purpose are the spectra of GRB-SN 2017iuk \citep[black;][]{izzo2019}, GRB-SN 1998bw \citep[grey;][]{patat2001}, and SN Ic-BL 2020bvc \citep[grey:][]{hiramatsu2020} at similar epochs (when available) on the bottom of each panel. The spectrum of SN 2020bvc is subtracted by an arbitrary power-law continuum to highlight the spectral features.
}
\label{fig:spec_early}
\end{figure*}

\begin{figure*}
\centering
\includegraphics[width=\columnwidth]{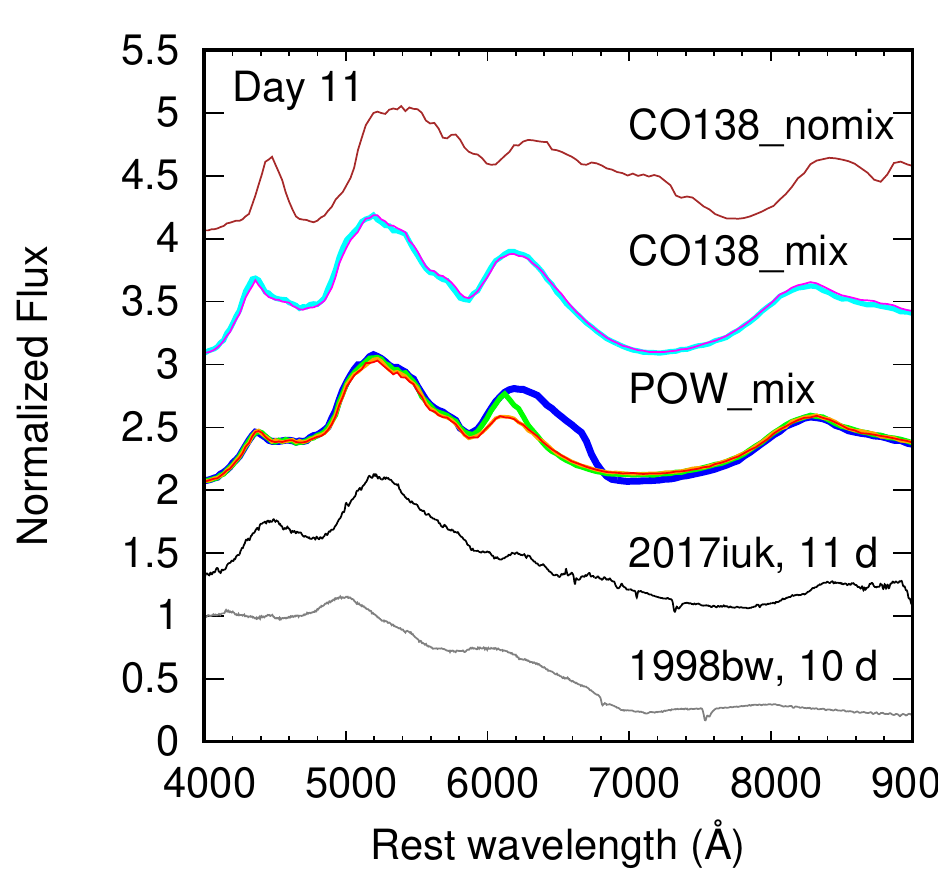}
\includegraphics[width=\columnwidth]{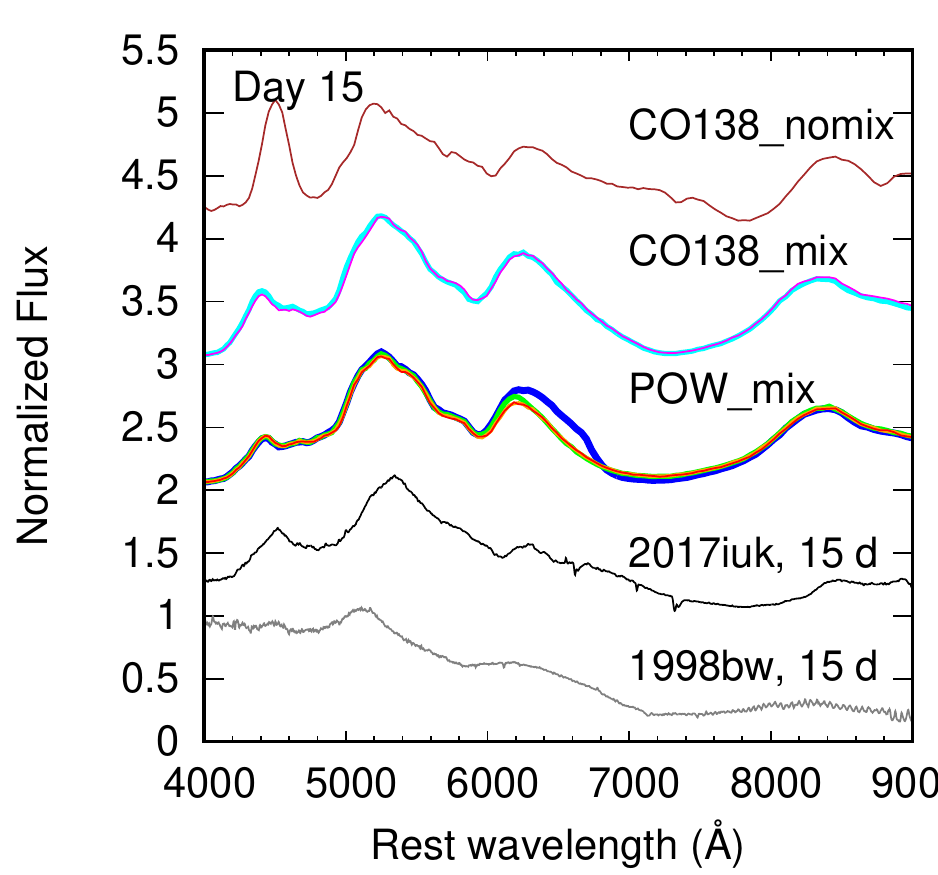}
\includegraphics[width=\columnwidth]{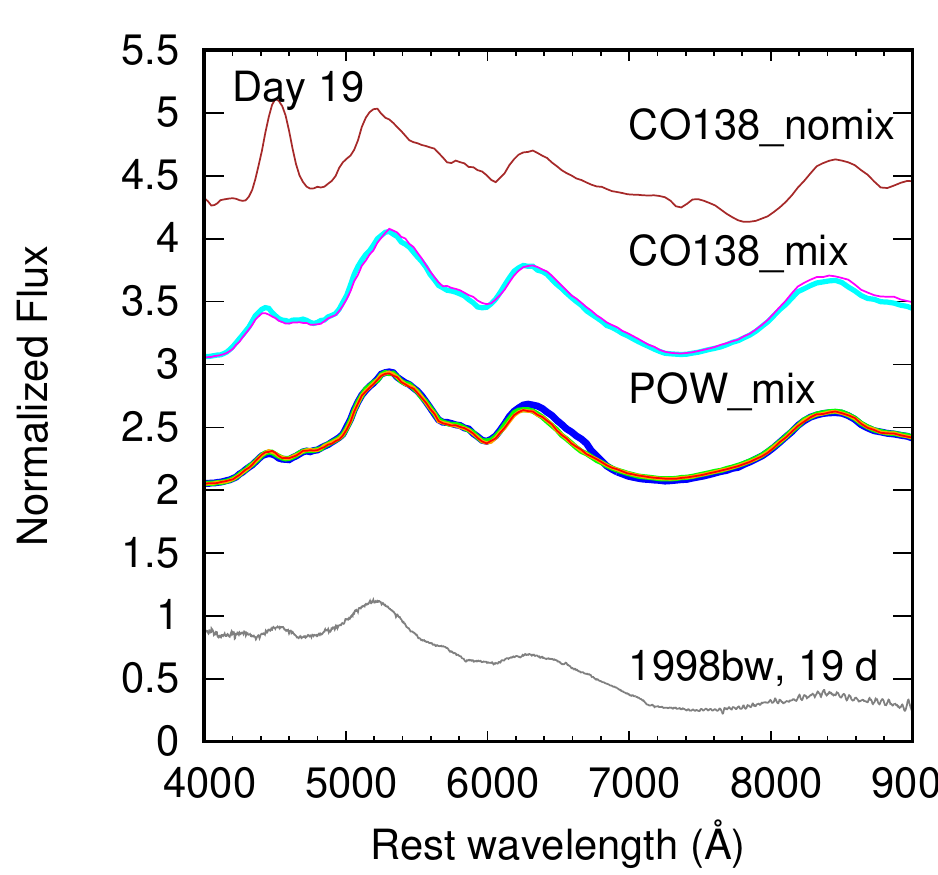}
\includegraphics[width=\columnwidth]{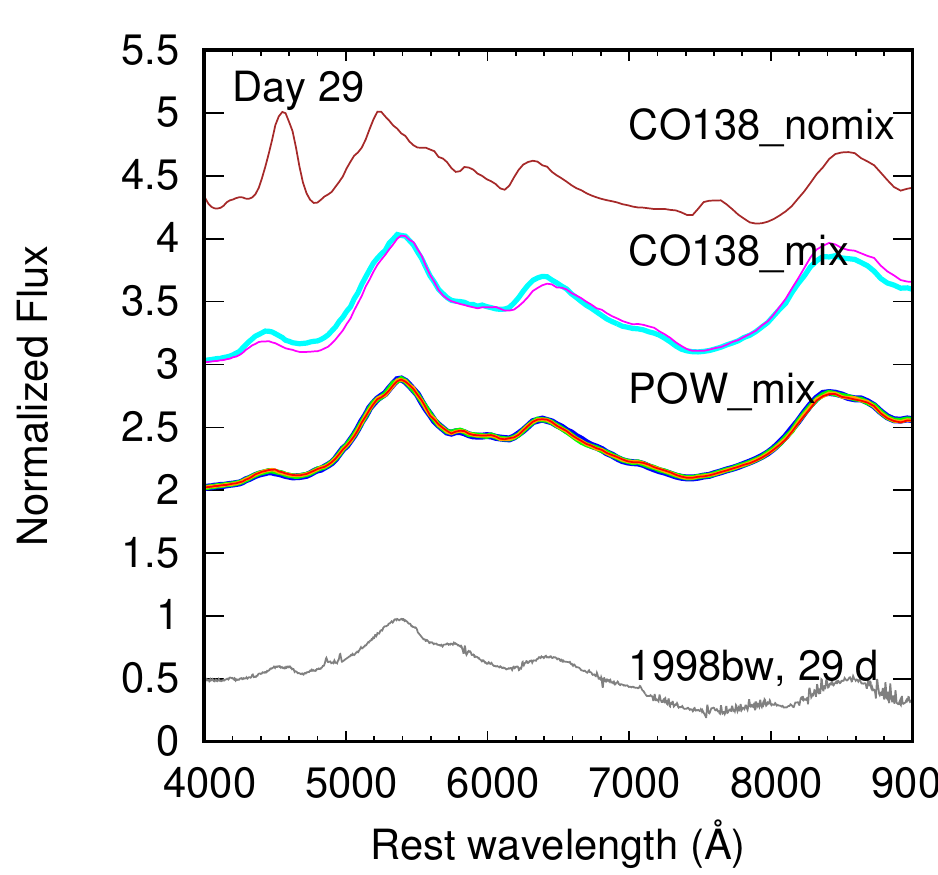}
\caption{The same as Figure \ref{fig:spec_early}, but for the maximum-light phases. 
}
\label{fig:spec_max}
\end{figure*}

Fig. \ref{fig:spec_early} shows the evolution of the synthetic spectra in the initial phase, on days 0.5, 1, 2, and 7, well before the maximum light\footnote{The synthetic spectra for the model CO138\_nomix are shown only on day 7 and thereafter. The model has essentially zero flux before day 7 (Fig. \ref{fig:lc}).}. While we do not intend to tune the models to fit the observational data of specific objects, we also show the spectra of GRB-SNe 1998bw \citep{patat2001}, 2007iuk \citep{izzo2019}, and SN Ic-BL 2020bvc \citep{hiramatsu2020} for a demonstration purpose\footnote{The spectra of SNe 1998bw and 2020bvc are downloaded from the WISeRep \citep{yaron2012}; \url{https://www.wiserep.org/}. The spectra of SN 2017iuk are available on GRBSpec \citep{deugarte2014}; \url{http://grbspec.eu/?index.php}.}. Fig. \ref{fig:spec_max} shows the spectral evolution covering the maximum phase. In the present work, we restrict ourselves up to about one month since the explosion for two reasons; (1) in the post-maximum phase, an additional dense core is required to explain both the light curve \citep{maeda2003b} and spectral properties (\citealt{maeda2006c}; see also \citealt{dessart2017}), and (2) the deviation from the LTE becomes progressively important toward a late phase. 

The chemical mixing manifests itself mainly in three effects. First, the emergence of the SN light becomes earlier for a more extensive $^{56}$Ni mixing, as already been discussed in many previous works \citep[e.g.,][see also Section 3.1]{nakamura2001,dessart2012,bersten2013,yoon2019}. The $^{56}$Ni mixing also affects the evolution of the photosphere, and therefore the photospheric velocity itself, e.g., at the maximum light \citep{dessart2016,moriya2020}. This is also seen in the spectral line velocities (Fig. \ref{fig:spec_max}). The other effect is the additional absorption in the `mixed' models especially in the blue, which is provided by metals, e.g., Fe II and Co II, in the outer layer (see the feature between 4,000 and 5,000 \AA\ in Fig. \ref{fig:spec_max}). 

The general behaviour in the time evolution, for different choice of the outermost ejecta velocity ($v_{\rm max}$), is clearly seen in the spectral evolution. Initially Models POW with different values of $v_{\rm max}$ show noticeable differences in the synthetic spectra as can be seen in the spectra on days 0.5, 1, and 2. On day 7, the models with $v_{\rm max} = 120,000$ and $100,000$ km s$^{-1}$ converge to show indistinguishable spectra, while the difference is still seen as compared to the other two models ($v_{\rm max} = 80,000$ and $60,000$ km s$^{-1}$). The similar behaviour is discerned between Models CO138 with $v_{\rm max} = 10,000$ and $80,000$ km s$^{-1}$, which show clear difference up to day 2, while the difference disappears on day 7. As time goes by, the difference between Models POW with $v_{\rm max} \ge 100,000$ and $80,000$ km s$^{-1}$ becomes smaller on day 11, and these two models become indistinguishable on day 15. Finally, their spectra merge into Model POW with $v_{\rm max} = 60,000$ km s$^{-1}$ on day 19, i.e., around the peak. 

The main difference between the model sequences CO138 and POW is the density structure at $\gsim 50,000$ km s$^{-1}$; the POW sequence has a flatter density structure and thus a higher density there than the CO138 sequence. It is seen in Fig. \ref{fig:spec_max} that the difference between these two sequences becomes progressively small toward the maximum light. If Models CO138 and POW with the same $v_{\rm max}$ are compared, the POW models show a more blue-shifted absorption feature (due to the line blending) than the CO138 models. 

The difference in the degree of the blueshift seen in the absorption as a function of $v_{\rm max}$ (as complemented by the density slope) explains the early-phase colour evolution for different models (Fig. \ref{fig:color} and Section 3.1). In the following, we focus on the model POW sequence, but essentially the same argument applies to the model CO138 sequence. Fig. \ref{fig:spec_early} shows that the synthetic spectra do not show noticeable difference in the $B$ band. The absorption feature in the models with large $v_{\rm max}$ starts overlapping the $V$-band bandpass, with the flux suppression in the $V$ band especially strong for Model POW\_mix\_12 in the first week. This is the reason why this model shows very blue $B-V$ colour in the initial phase, followed by Model POW\_mix\_10. The $R$-band flux is more easily affected by the absorption. Therefore, models with large $v_{\rm max}$ suffered from the flux suppression both in the $V$ and $R$ bands, while decreasing $v_{\rm max}$ leads to the suppression only in the $R$ band. This explains the non-monotonic behaviour as a function of $v_{\rm max}$ seen in the $V-R$ colour. Note that the difference here is {\em not} due to difference in the photospheric temperature -- it is seen in Figs. \ref{fig:spec_early} and \ref{fig:spec_max} that Models POW\_mix (and Models CO138\_mix) have essentially the same continuum colour irrespective of $v_{\rm max}$\footnote{Note that the photospheric velocities derived for SN 2017iuk were $59,000$ and $53,000$ km s$^{-1}$ on days 1 and 2, respectively \citep{izzo2019}. These are already below the minimum value of $v_{\rm max}$ (60,000 km s$^{-1}$) examined in the present work.}; except for the wavelength ranges that experience the high-velocity absorption associated with the materials well above the photosphere, the synthetic spectra for the models with different values of $v_{\rm max}$ are overlapping (e.g., at $\lsim 4,500$\AA\ and $\gsim 7,000$\AA) -- if the photospheric temperature is different, the overal spectral slope must be different. This effect of $v_{\rm max}$ on the colour is thus different from that by the $^{56}$Ni mixing, later of which changes the photospheric velocity and temperature \citep{yoon2019}. 

The above considerations show a potential power of the spectroscopic observations (as well as the less-direct photometric observations) to diagnose the properties of the outermost layer, which is otherwise difficult. The spectra in the first few days provide dramatic difference in the degree of blueshift in the (blended) absorption feature at 5,000-7,000\AA, depending on $v_{\rm max}$. This dependence on $v_{\rm max}$ is to some extent coupled with the density structure, since the larger density in the outermost layer (i.e., Models POW) tend to show a larger blueshift in the absorption feature than the lower-density models (i.e., Models CO138), for given $v_{\rm max}$ (i.e., the cyan line vs. the green line, or the magenta line vs. the orange line, in Fig. \ref{fig:spec_early}). The time evolution is however somewhat different (e.g., Fig. \ref{fig:color} for the $V-R$ colour evolution; Figs. \ref{fig:spec_early} and \ref{fig:spec_max} for the longer appearance of the additional absorption in Models POW than CO138 for same $v_{\rm max}$), and the detailed time-series modeling could help solve the degeneracy. 

To demonstrate the power of the very early-phase spectra to diagnose the ejecta properties, we compare the synthetic model spectra to the observational ones of GRB-SNe 2017iuk and 1998bw, and SN Ic-BL 2020bvc, with the caveat that the present models are not tuned to reproduce observational properties of specific objects. The intensive spectral coverage in the initial, rising phase is available for GRB-SN 2017iuk \citep{izzo2019}. Fig. \ref{fig:spec_early} shows that Models POW\_mix\_10 and POW\_mix\_12 capture key spectral properties of SN 2017iuk. On day $\sim 1$, a broad absorption feature is seen in $\sim 5,200-6,400$\AA\ with the minimum at $\sim 5,800$\AA\ in the observation. The models (both POW and CO138) with $v_{\rm max} \lsim 80,000$ km s$^{-1}$ instead show an emission-like feature peaking at $\sim 6,000$\AA, contrary to the observational spectrum. With $v_{\rm max} \sim 100,000$\ km s$^{-1}$, this peak starts being suppressed by a blue-shifted absorption component originally in the redder wavelength, creating instead the absorption feature in the same wavelength range. This property persists on day 2, where again the models with $v_{\rm max} \gsim 100,000$ km s$^{-1}$ are favored. The same effect is seen in the model spectra on days 7 and 11; the peak at $\sim 6,200$\AA\ seen in the models with the smaller $v_{\rm max}$ starts being overlapped by the absorption for Model POW\_mix\_8, and it is totally smeared out to produce a featureless continuum-like spectra in $\sim 5,500 - 8,000$\AA\ in Models POW\_mix\_10 and POW\_mix\_12. The latter is consistent with the observational spectrum. Model CO138\_mix\_10 fails to explain this feature due to the lower density than Model POW\_mix\_10 in the line-formation region. All the models start showing the peak at $\sim 6,200$\AA\ on day 15, when the corresponding emission feature is discerned in the observed spectrum. In summary, despite various simplifications in the model construction, the power-law density structure (with the index of $\sim -6$ in the velocity space) extending to $\gsim 100,000$ km s$^{-1}$, with substantial mixing in the composition structure including $^{56}$Ni, reproduces the spectral evolution of SN 2017iuk reasonably well. We thus confirm the main conclusions of \citet{izzo2019}, emphasizing that the formation of the photosphere is treated in a self-consistent (first-principle) manner in the present work, rather than the parameterized treatment of the photosphere adopted in many spectral-synthesis studies including the models presented in \citet{izzo2019}. 

As emphasized above, this comparison is for a demonstration purpose, given that we have not tuned the models to individual objects. For example, on day 1, the main absorption feature discussed above has a wide absorption feature up to $\sim 7,500$\AA\ in the models, while the observed spectrum of SN 2017iuk shows a peak at $\sim 6,500$ km s$^{-1}$. In addition, the observed spectrum does not show a clear peak at $\sim 4,800$\AA, which is seen as a strong peak in the model spectra. These may be affected by the detailed density and composition structures more sensitively than the absorption structures (including the possibility of non-spherical geometry); the absorption mainly probes the information along the line-of-sight toward the photosphere, while the emitting region is generally more extended. Further, in the earliest phase of SN 2017iuk, there is an uncertainty in the subtraction of the GRB afterglow component, which may artificially create or smooth peak structures while the absorption features are less affected by such subtraction. As such, we focus on the absorption features in the present work, but emphasize that more detailed study with finer/wider model grids is required for detailed spectral study of individual objects \citep[e.g.,][]{izzo2019}. 

SN Ic-BL 2020bvc was not associated with a GRB, but it was discovered soon after the explosion thanks to the recent development of high-cadence optical surveys. The follow-up observations revealed that it belongs to an extreme end of SNe Ic-BL, with many properties shared with GRB-SNe, including the high-expansion velocity as well as strong radio and X-ray emissions \citep{ho2020,izzo2020,rho2021}. SN 2020bvc has been thus suggested to be driven by a powerful central engine creating a sub-relativistic ejecta component, either by a chocked jet \citep{ho2020} or an off-axis jet \citep{izzo2020}. Very high-velocity absorption features reaching to $\sim 60,000 - 70,000$ km s$^{-1}$ are seen in its spectrum taken on day 1.5, which shares the property seen in SN 2017iuk at a similar epoch. It then evolved to show similar maximum-light spectra with SN 1998bw. Fig. \ref{fig:spec_early} shows the spectrum on day 1.5. SN 2020bvc showed a blue continuum on day 1.5; to focus on the absorption/emission features, its spectrum shown in Fig. \ref{fig:spec_early} is subtracted by an arbitrarily-defined power-law continuum and then the flux is normalized. The difference to SN 2017iuk is striking, which may be mainly attributed to the difference in the outermost velocities; the spectrum resembles the models with $v_{\rm max} = 60,000$ or $80,000$ km s$^{-1}$ (i.e., CO138\_mix\_8, POW\_mix\_6, and POW\_mix\_8), confirming the claim by \citet{ho2020,izzo2020,rho2021}. SN 2020bvc demonstrates three key issues; (1) the very-early spectra of GRB-SNe and SNe Ic-BL can be used to constrain the outermost ejecta properties, (2) there could indeed be a diversity in the outermost ejecta properties among GRB-SNe and SN Ic-BL, and they might further be linked to their association or non-association with GRBs, and (3) the high-cadence optical surveys now provide an excellent opportunity to perform systematic investigation of the outermost ejecta properties, even without an associated GRB. 

A caveat here on the comparison between the present models and the spectrum of SN 2020bvc is that the models presented here take into account only the $^{56}$Ni/Co heating, while SN 2020bvc probably exhibited a cooling emission either from a jet-heated cocoon or a confined circumstellar matter (CSM) hit by the sub-relativistic ejecta \citep[][]{izzo2020,ho2020,jin2021,rho2021}; we plan to extend the model to include the cooling emission in the early-phase spectral formation \citep[see][for the light curve simulations]{suzuki2019b,suzuki2021}. Indeed, the present investigation may already place a potential constraint on the underlying mechanism for the early, bright and rapidly declining emission. \citet{jin2021} and \citet{rho2021} showed that the early light curve evolution of SN 2020bvc can be explained by the cooling emission resulting from the interaction between the SN ejecta and a dense and confined CSM, with the mass of $\sim 0.05-0.3 M_\odot$. The expectation from this scenario is that the high-velocity ejecta should be decelerated, and thus the formation of the high-velocity blueshifted absorption may indeed be suppressed. Taken the model POW\_mix\_8 as an example (with $v_{\rm max} = 80,000$ km s$^{-1}$), Fig. \ref{fig:velocity} shows that the reverse shock will decelerate the outermost ejecta to $\sim 60,000$, $40,000$, and $30,000$ km s$^{-1}$ for the CSM masses of $\sim 0.05$, $0.1$, and $0.3 M_\odot$, respectively, according to the rough estimate that the swept-up mass of the CSM and that of the ejecta are comparable. Given that the high-velocity absorption features reaching to $\sim 60,000 - 70,000$ km s$^{-1}$ are seen in its earliest spectrum, the consideration here places a strong constraint on the SN-CSM interaction scenario; the scenario is generally disfavored, with only the least massive CSM case ($\lsim 0.05 M_\odot$) still surviving as a possibility. We will investigate the origin of the early emission of SN 2020bvc in a forthcoming paper (K. Maeda, in prep.).

The same level of comparison to SN 1998bw is not possible as the early-phase spectra are observationally missing. We however note that SNe 2017iuk and 1998bw show overall similarities in the maximum-phase spectra when spectra of both SNe are available, despite the difference of $\sim 1$ magnitude at the peak. SN 1998bw shows larger degree of blueshift in overall spectral shape than SN 2017iuk, which probably reflects some intrinsic difference in the ejecta kinematics \citep{izzo2019}. The viewing angle effect is probably not strong in the maximum phase for models constructed for GRB-SNe where the SN component is quasi-spherical, and both SNe are probably viewed on-axis \citep{maeda2006,tanaka2007,rapoport2012,barnes2018,shanker2021}, and it would not explain the difference between the two SNe. SN 1998bw is bluer than the present models, which might be explained by a low metallicity \citep{rapoport2012} or by a less substantial mixing than in the `mixed' models. In any case, the overall spectral properties are well explained by the present models. 

%; we emphasize again that most of the previous spectral fitting study for GRB-SNe assumed a sharp-photosphere and its properties (velocity and temperature) as input parameters, and if the photosphere is consistent with `what is assumed as an input parameter' is another story. The present models automatically compute the photosphere formation process based on the first-principle calculation without tuning any parameter once the input ejecta model are specified, and thus the success of the present models to explain the key features seen in the prototypical SN 1998bw is an important step forward to understand the nature of the central engine behind GRB-SNe. 

\section{Concluding remarks}\label{sec:conclusions}

In the present paper, we demonstrate how the ejecta properties of GRB-SNe and engine-driven SNe Ic can be constrained through the comparison between the synthetic observables (both in photometry and spectroscopy) and the observed data, especially focusing on the initial, rising phase soon after the explosion. Based on insights obtained through a series of the hydrodynamic and radiation-hydrodynamic simulations of the engine-driven SNe \citep{suzuki2016,suzuki2017,suzuki2019,suzuki2021,suzuki2022}, we investigate the effects of the composition mixing, density structure, and the maximum ejecta velocity. Out findings can be summarized as follows: 
\begin{enumerate}
    \item The different degree of the composition mixing, especially of $^{56}$Ni, changes the rising time and the colour evolution substantially. The effect is also seen in synthetic spectra, where the line velocities are also affected, with the mixing leads to the higher velocities.
    \item The density slope in the outermost layer affects the initial spectral evolution, where the flatter density distribution, as expected from the hydrodynamic behaviour induced by the central engine, leads to high velocities in the spectral features. 
    \item The outermost ejecta velocity can be strongly constrained from the spectroscopic data in the initial phase ($\lsim$ a week since the explosion), in a way that the higher velocity leads to broader spectral features resulting in featureless spectra. 
    \item The dependence of the colour evolution on the parameters (e.g., the maximum ejecta velocity) is not necessarily monotonic. The characteristic colour evolution for different models can be understood from the spectral formation processes affecting fluxes in different bandpasses. 
    \item The effects of the density structure and the maximum velocity are seen clearly in the infant phase (within a week since the explosion in the model sequence considered in the present work). The synthetic observables however eventually converge toward the maximum phase, leaving the earliest phase observation critical to diagnose the ejecta properties. 
\end{enumerate}

We compare the synthetic spectra with the observed spectral sequences of SNe 1998bw, 2017iuk, and 2020bvc. While our models are simplified (e.g., assuming spherically symmetric structure; see below), and not tuned to fit the data (e.g., assuming a complete mixing and investigating only a fixed set of the ejecta mass and kinetic energy), the present models reproduce the observed spectra reasonably well. The characteristic featureless spectra due to the blending of individual broad lines are well explained. The spectra in the first week are not available for SN 1998bw, and not much constraint can be placed on the density structure and maximum ejecta velocity. On the other hand, the intensive spectral sequence in the initial phase available for SN 2017iuk provides powerful diagnostics on the ejecta properties. The power-law density structure as combined with the maximum ejecta velocity of $\gsim 100,000$ km s$^{-1}$ provides good reproduction of the spectral features and their evolution. The similar analysis is possible for non-GRB (or off-axis GRB) SN Ic BL 2020bvc, which also favors the existence of the high-velocity ejecta but with somewhat lower maximum velocity, i.e., $60,000 - 80,000$ km s$^{-1}$. These analyses highlight the power of the infant-phase spectroscopic observations of GRB-SNe and SNe Ic-BL. 

One important caveat in the present investigation is that we have assumed the spherically symmetric ejecta. This is indeed the case for most of the previous works; based on the success of such models and the accumulated experiences obtained there, we believe that the characteristic ejecta properties, i.e., the density and composition structures as well as the maximum velocity, are reasonably well constrained within this framework. Indeed, these `spherical' properties may be regarded to represent those along the line of sight (thanks to the large optical depth in the initial phase up until the maximum phase). For example, a series of the modeling activities for SN 1998bw using a jet-like explosion model suggests that this argument is largely justified; the required energy may be reduced by a factor of a few as compared to the spherical model, but otherwise many conclusions based on the spherical models would not change \citep{maeda2006,maeda2006b,maeda2006c,tanaka2007,rapoport2012,barnes2018}, provided that the key multi-dimensional effects, i.e., the change in the density structure and mixing in the composition structure, are taken into account. 

GRB observations are triggered by the $\gamma$-ray photons in an unbiased way, and thus a search for an emerging SN component in optical wavelengths is possible starting just after the detection of the GRB. When the SN component is detectable is dependent on the competition between the decreasing contribution of the GRB afterglow and the increasing contribution of the SN component toward the maximum light. Therefore the clear detection of the SN component in the first few days after the explosion requires some specific conditions, e.g., close distance and weak GRB afterglow including possible off-axis events. So far, GRB 161219B/SN 2016jca and GRB 171205A/SN 2017iuk have provided such opportunities, with the SN spectra showing broad features available already 2 days (SN 2016jca) and even 1 day (SN 2017iuk) after the explosion. While further increasing the sample depends on luck, it is important not to miss the chance to obtain such irreplaceable data set for the next (rare) nearby GRB. 

Indeed, the strategy may work even better for SNe Ic-BL without an associated GRB. Thanks to the recent advance of the new-generation, unbiased and wide-field optical surveys, e.g., Pan-STARRS \citep[Panoramic Survey Telescope And Rapid Response System;][]{kaiser2002}, ATLAS \citep[Asteroid Terrestrial-impact Last Alert System;][]{tonry2018} and ZTF \citep[Zwicky Transient Facility;][]{bellm2019,masci2019}, nearby SNe are routinely discovered in the infant phase just after the explosion. For a survey limiting magnitude of $\sim 20$ mag, SNe within the distance of $\sim 100$ Mpc can be discovered at the absolute magnitude of $\sim -15$ mag; taking the light curve of SN 1998bw as a template, this corresponds to the magnitude within a day of the explosion, which allows a prompt spectroscopic observations to obtain the spectra within a few days, or even within a day, of the explosion. Within this distance, $\sim 100 - 200$ CCSNe are expected to be discovered per year. Given the relative frequency of SNe Ic-BL being $\sim 1- 2$\% among CCSNe \citep{grauer2017}, we may expect that infant spectra can be obtained for a few SNe Ic-BL per year. The model predictions in the present work can be applied to such data, which will hopefully form essential background to investigate whether and how the ejecta properties are different between GRB-SNe and non-GRB SNe Ic-BL. While being rare, such investigation will particularly important for non-GRB `engine-driven' SNe Ic-BL, such as SNe 2009bb \citep{soderberg2010,pignata2011}, 2012ap \citep{margtti2014,milisavljevic2015}, 2014ad \citep{sahu2018,kwok2022}, and 2020bvc \citep{ho2020,izzo2020,rho2021}.
 
\section*{Acknowledgements}
The authors thank the anonymous referee for her/his constructive and insightful comments. K.M. acknowledges support from the Japan Society for the Promotion of Science (JSPS) KAKENHI grant JP18H05223, JP20H00174, and JP20H04737. A.S. acknowledges support from the JSPS KAKENHI grant JP19K14770 and JP22K03690. Numerical computations were carried out on Cray XC50 at Center for Computational Astrophysics, National Astronomical Observatory of Japan. We also used the Yukawa Institute Computer Facility. Some data presented in this work are obtained from WISeREP (\url{https://www.wiserep.org}) and GRBspec database (\url{http://grbspec.eu/?index.php}).

%%%%%%%%%%%%%%%%%%%%%%%%%%%%%%%%%%%%%%%%%%%%%%%%%%
\section*{Data Availability}
The simulated light curves and spectra are available upon request to K.M.



\bibliographystyle{mnras}
\bibliography{grbsn} % if your bibtex file is called example.bib

%\begin{thebibliography}{99}
%\end{thebibliography}

%%%%%%%%%%%%%%%%%%%%%%%%%%%%%%%%%%%%%%%%%%%%%%%%%%

%%%%%%%%%%%%%%%%% APPENDICES %%%%%%%%%%%%%%%%%%%%%

%\appendix

%%%%%%%%%%%%%%%%%%%%%%%%%%%%%%%%%%%%%%%%%%%%%%%%%%

% Don't change these lines
\bsp	% typesetting comment
\label{lastpage}
\end{document}